\documentclass[]{JHEP3}

\JHEPspecialurl{http://jhep.sissa.it/JOURNAL/JHEP3.tar.gz}

\usepackage{graphicx}
\usepackage{amsmath}
\usepackage{epsfig,multicol,bbm}

\newcommand{\be}{\begin{equation}}
\newcommand{\ee}{\end{equation}}
\newcommand{\bea}{\begin{eqnarray}}
\newcommand{\eea}{\end{eqnarray}}

\newcommand{\ti}{\times}
\newcommand{\half}{\frac{1}{2}}
\newcommand{\mc}{\mathcal}

\newcommand{\beqa}{\begin{eqnarray}}
\newcommand{\eeqa}{\end{eqnarray}}

\newcommand{\nn}{\nonumber}

\newcommand\fverb{\setbox\fverbbox=\hbox\bgroup\verb}
\newcommand\fverbdo{\egroup\medskip\noindent%
			\fbox{\unhbox\fverbbox}\ }
\newcommand\fverbit{\egroup\item[\fbox{\unhbox\fverbbox}]}
\newbox\fverbbox

\title{Moduli-Induced Vacuum Destabilisation}

\author{Joseph P. Conlon, Francisco G. Pedro\\
	Rudolf Peierls Centre for Theoretical Physics, University of Oxford, 1 Keble Road, Oxford, OX1 3NP, UK\\
	E-mail: \email{j.conlon1@physics.ox.ac.uk}, \email{f.pedro1@physics.ox.ac.uk}}

\preprint{OUTP-10-28P}	

\abstract{We look for ways to destabilise the vacuum. We describe how dense matter environments
source a contribution to moduli potentials and analyse the conditions required
to initiate either decompactification or a local shift in moduli vevs.
We consider astrophysical objects such as neutron stars as well as cosmological and black hole singularities.
 Regrettably neutron stars cannot destabilise realistic Planck coupled moduli, which would require objects
 many orders of magnitude denser. However gravitational collapse, either in matter-dominated universes or in black hole formation,
 inevitably leads to a destabilisation of the compact volume causing a super-inflationary expansion of the extra dimensions.
}

\begin{document}

\section{Introduction}

String theory has no free parameters. All coupling constants are instead determined as vacuum expectation values of scalar fields - moduli.
The values these scalar fields take are determined by the moduli potential, and these values determine the parameters of the Standard Model
and through them the masses, couplings and interactions of all known particles.

Moduli potentials have many ingredients, and much work has been done on constructing potentials that stabilise the moduli in phenomenologically
attractive fashions. However moduli vevs are environmental and so there is no reason in principle why they should stabilise at the same values
at all regions in space and time - indeed, we should expect the converse. In this article we therefore look at ways of destabilising moduli
from their vacuum values.

The basic mechanism we investigate is simple: as moduli source the couplings of Standard Model fields, any form of energy density represents
a source for the moduli potential. If the local energy density is sufficiently large - and large energy densities are realised
both within neutron stars and in the context of cosmological singularities - then this gives a contribution to the moduli potential
that can destabilise the modulus vev from its minimum.

Destabilised moduli vevs may be hard to achieve but the payoff if it can be done is large:
\begin{itemize}

\item
Small changes in moduli vevs would give a continuous deformation away from the Standard Model, with associated small shifts
in particle masses and couplings.

\item
Any region in which moduli vevs are altered can catalyse exotic processes
that are forbidden or highly suppressed within the Standard Model. A simple example is
that of proton decay. Within the Standard Model the proton can decay via electroweak sphaleron processes. However these are
nonperturbative and are suppressed by $\mc{A} \sim \exp (-S) \sim e^{-\frac{ 8 \pi^2}{g_{SU(2)}^2}}$ and the resulting proton lifetime
is of order $10^{\mc{N}}$ years with $\mc{N} \sim 100$.
Any local region in which the $SU(2)$ coupling were substantially stronger could therefore catalyse the decay of protons that
enter it.

\item
Large shifts in moduli can potentially provide windows into entirely different vacua of the underlying theory,
which could perhaps be realisable locally as stable solitonic objects.

\end{itemize}

For these reasons we think it worthwhile to investigate the possibility of destabilising moduli in various contexts. This discussion
can be made concrete by recent developments in moduli stabilisation, as it is difficult to discuss the chances of destabilising moduli
without concrete and well-motivated potentials that first stabilise moduli.

For other related work, see for example \cite{0309300, 0408464, Green:2006nv, 10053735}.

\section{The effective field theory}

We study the possibility and the consequences of moduli/matter interaction within the framework of four dimensional supergravity. Starting from the Type IIB action in 10 dimension, upon dimensional reduction in the large volume limit one finds that the action for the moduli sector is
\be
S=\int d^{4}x\sqrt{-g}\left(G_{i\bar{j}}\partial_{\mu}\phi^{i}\partial^{\mu}\phi^{\bar{j}}-V(\phi,\bar{\phi})\right),
\ee
where  $G_{i\bar{j}}=\partial_{i}\partial_{\bar{j}} K$ is the K\"ahler metric and $\phi_{i}$ are the moduli fields of the theory, i.e. the dilaton, complex structure and K\"ahler moduli. The F-term potential takes its usual form
\be
V(\phi,\bar{\phi})=e^K (G^{i \bar{j}} D_i W D_{\bar{j}}\bar{W} -3 |W|^2),
\label{eq:V}
\ee
where $D_{i}K=\partial_{i}K+K\partial_{i}W$.

We will work within the LARGE volume models \cite{hepth0502058}, which allow for the stabilisation of all the moduli while generating a nonsupersymmetric AdS minimum at exponentially large volumes.
We briefly review the properties of these models and the masses of the moduli in them.
These models are characterised by the inclusion of $\alpha'^{3}$ corrections to the K\"ahler potential \cite{BBHL}:
\be
K=-2\ln \left( \mathcal{V} +\frac{\xi(\frac{S+\overline{S}}{2})^{3/2}}{2} \right) + K_{cs}+K_{S},
\label{eq:K}
\ee
as well as  nonperturbative effects in the superpotential $W$ \cite{kklt}:
\be
W=W_{0}+ \sum_{i}A_{i} e^{i a_{i} \rho_{i}}.
\label{eq:W}
\ee

It was shown in \cite{hepth0502058} that these models have AdS minima with broken SUSY for geometries in which one K\"ahler modulus controls the overall volume while the remaining K\"ahler moduli describe collapsible four cycles. This so called `Swiss cheese' geometry is quite generic and various explicit cases have been described in \cite{0404257, 07113389, 08114599}. In this work we will focus on the simplest realisation of this geometry, the two modulus model defined by a degree 18  hypersurface in the weighted projective space $\mathbb{P}^{4}_{[1,1,1,6,9]}$. The volume for this orientifold is given by
\be
\mathcal{V}=\frac{1}{9\sqrt{2}}(\tau_{b}^{3/2}-\tau_{s}^{3/2}),
\ee
where the modulus $\tau_{b}$ parametrizes the volume of the extradimensional manifold and $\tau_{s}$ the size of a `small' four cycle in the manifold. We note that the results found here should remain qualitatively the same for more complicated manifolds provided they are of the Swiss cheese type.

The scalar potential, Eq. (\ref{eq:V}), for the LARGE volume models can be computed by taking the K\"ahler potential and the superpotential, Eqs. (\ref{eq:K}) and (\ref{eq:W}). One can choose to  stabilise the axio-dilaton and the complex structure moduli in a SUSY preserving way, $DW=0$ , by turning on fluxes in the extra dimensions. This will cause their contribution to the scalar potential to vanish. For the two moduli model we are considering the scalar potential can be written as:
\be
V=C_{1} \frac{\sqrt{\tau_{s}} e^{-2 a \tau_{s}}}{\mathcal{V}}-C_{2} \frac{\tau_{s} e^{-a \tau_{s}}}{\mathcal{V}^{2}}+\frac{C_3}{\mathcal{V}^{3}},
\label{eq:VLVS}
\ee
where
\bea
C_{1}=\frac{8}{3}\lambda a^{2} |A|^{2},\\
C_{2}=4 |A W| a, \\
C_{3}= \frac{3}{4}\frac{|W|^{2} \xi}{g_{s}^{3/2}}.
\eea

This potential has a LARGE volume AdS minimum at
\bea
\tau_{s}^{3/2}= \frac{\lambda \xi}{2 g_s^{3/2}} \left(1-\frac{1}{2a\tau_{s}}\right),\\
\mathcal{V}=\frac{3 |W| \sqrt{\tau_{s}}e^{a \tau_{s}}}{\lambda a |A|}\left(1-\frac{3}{4a\tau_{s}}\right).
\eea

\subsection{Moduli spectrum}
Given that the moduli have been stabilised at the minimum of the potential one can compute their masses:
\be
m_{i}^{2}=G^{i \bar{i}}\frac{\partial^{2} V}{\partial \tau_{i}^{2}},
\ee
where in the particular case under consideration $i\in \{s,b\}$.
Noting that the leading terms in the volume expansion of the inverse K\"ahler metric are
\bea
G^{s \bar{s}}\propto\mathcal{V},\\
G^{b \bar{b}}\propto\mathcal{V}^{4/3},
\eea
once we compute the second derivatives of the scalar potential Eq. (\ref{eq:VLVS}) and restore the factors of $M_{P}$ we find
\bea
m_{s} \approx \frac{M_{P}}{\mathcal{V}},\\
m_{b} \approx \frac{M_{P}}{\mathcal{V}^{3/2}}.
\label{eq:masses}
\eea

In the LARGE volume minimum one finds that the small modulus is much heavier that the large modulus. One can therefore consider that it decouples from the theory, and take $\tau_{b}$ to be the only dynamical variable in the problem. This corresponds to using
\be
e^{a \tau_{s}}=\mathcal{V} (1 + \hbox{subleading terms}),
\label{eq:eatau}
\ee
to eliminate $\tau_{s}$ dependence. We will later apply this to the scalar potential, Eq. (\ref{eq:VLVS}), but first let us formulate the problem in terms of canonically normalised fields.

\subsection{Canonical normalisation of the volume modulus}

As discussed in the previous section, the moduli associated with the volume of the `small' four cycle gets, through moduli stabilisation, a very large mass and can therefore be integrated out. What remains is the theory of a single scalar field. The kinetic part of the Lagrangian is \footnote{recall that $G_{i\bar{j}}=\frac{\partial^{2}}{\partial T_{i}\partial T_{\bar{j}}} K$ and $T_{i}= \tau_{i}+ib_{i}$}
\be
L_{K}=G_{b \bar{b}}\partial_{\mu}T_{b}\partial^{\mu}T_{\bar{b}}=\frac{3}{4\tau_{b}^2}\partial_{\mu}\tau_{b}\partial^{\mu}\tau_{b}.
\ee
For convenience we will work with the canonically normalised field $\Phi$, defined by
\be
\frac{3}{4\tau_{b}^2}\partial_{\mu}\tau_{b}\partial^{\mu}\tau_{b}=\frac{1}{2}(\partial\Phi)^{2},
\ee
so we find
\be
\Phi=\sqrt{\frac{3}{2}}\ln\tau_{b}=\sqrt{\frac{2}{3}}\ln \mathcal{V}.
\label{eq:Phi}
\ee

\begin{figure}[h]
\begin{center}
\includegraphics[width=0.5\textwidth]{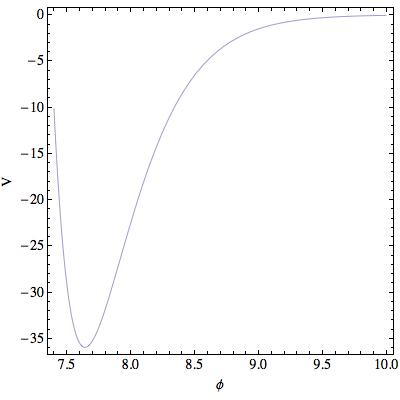}
\caption{Potential (multiplied by $10^{15}$) for the canonically normalised volume modulus for $\alpha=0.05$.}
\label{fig:LVS}
\end{center}
\end{figure}

Using Eqs. (\ref{eq:eatau}) and (\ref{eq:Phi}) one rewrites the potential, Eq. (\ref{eq:VLVS}), as
\be
V=(1-\alpha \Phi^{3/2}) e^{-\sqrt{\frac{27}{2}}\Phi},
\label{eq:VPhi}
\ee
where we have ignored factors of order unity (note this requires including the subleading terms of eq. (\ref{eq:eatau})). In Fig. \ref{fig:LVS} we plot the potential for the canonically normalised volume modulus, Eq. (\ref{eq:VPhi}), which exhibits the characteristic AdS minimum at exponentially large volume.

\begin{figure}[h]
\begin{center}
\includegraphics[width=0.5\textwidth]{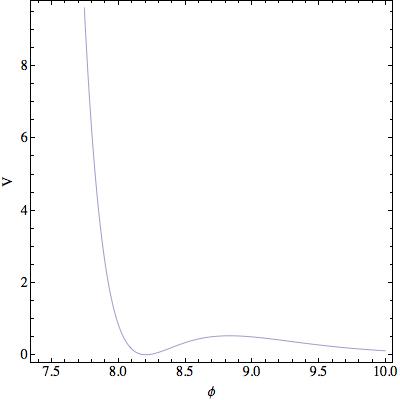}
\caption{The full potential (multiplied by $10^{15}$) for the canonically normalised volume modulus for $\alpha=0.05$.}
\label{fig:LVSUP}
\end{center}
\end{figure}

In order to obtain a phenomenologically viable model, it is necessary to lift the AdS minimum to dS or Minkowski, without spoiling the stabilisation of the K\"ahler moduli of the theory. The procedure proposed in \cite{kklt} is to add a $\overline{D3}$ brane which generates a term in the potential of the form:
\be
V_{\overline{D3}}\propto \frac{1}{\mathcal{V}^{2}}=e^{-\sqrt{6}\Phi}.
\ee
The full potential then takes the form
\be
V=(1- \alpha\Phi^{3/2}) e^{-\sqrt{\frac{27}{2}}\Phi}+ \epsilon e^{-\sqrt{6}\Phi}.
\label{eq:Vfull}
\ee
By tuning $\epsilon$ one can then find a dS or Minkowski minimum with all moduli stabilised. The minimum will lie in the same region as the initial AdS as illustrated in Fig. \ref{fig:LVSUP}.

\section{Moduli/matter interaction}

The goal of this work is to investigate the interaction between matter and moduli fields. In particular,
we want to know whether very large energy densities can destabilise the moduli or create bubbles of different vacua,
 and if so what will be the observable consequences.

Let us first enumerate some general features of this interaction. First, the volume modulus couples to everything. This is
primarily because it sets overall scales and enters all dimensionful quantities. Secondly, the volume modulus is also
the lightest modulus. This implies that it is the most appropriate field to consider the interactions of, as it should be easiest
to destabilise. Finally, in order to try and destabilise the field we want the densest possible regions in order
to have maximal effect on the
potential.

Natural candidates to consider are neutron stars. These are very dense, gravitationally bound systems made up mostly of neutrons. We start by noting that the energy density of a neutron star is:
\be
\rho_{NS}\approx \Lambda_{QCD}^{4}.
\ee
Due to gauge coupling running this term will be moduli dependent and therefore it will contribute to the moduli potential, Eq. (\ref{eq:Vfull}).
Let us briefly illustrate how this occurs. Recall that the QCD $\beta-$function  is
\be
\beta(g(\mu))=-\frac{9}{16 \pi^{2}}g(\mu)^{3},
\label{eq:betaQCD}
\ee
where as usual
$
\beta(g(\mu))\equiv\mu\frac{d g}{d\mu}.
$
Substituting the definition of the $\beta-$function into Eq. (\ref{eq:betaQCD}) and integrating one finds that
\be
\ln(\mu'/\mu'')=-\frac{8 \pi^{2}}{9}\left(\frac{1}{g^{2}(\mu')}-\frac{1}{g^{2}(\mu'')}\right),
\ee
and setting $g(\mu')\rightarrow \infty$, $\mu'\equiv\Lambda_{QCD}$ yields
\be
\Lambda_{QCD}=\mu'' e^{-\frac{8 \pi^{2}}{9 g^{2}(\mu'')}}.
\ee
Assuming that the coupling starts to run from the string scale $M_s$, we set
\bea
\mu''=M_{s}=\frac{M_{P}}{\sqrt{\mathcal{V}} }, \\
g^{2}(\mu'')=g_{YM}^{2},
\eea
to find
\be
\label{proton}
\Lambda_{QCD}=\frac{M_{P}}{\sqrt{\mathcal{V}} } e^{-\frac{8 \pi^{2}}{9 g_{YM}^{2}}}.
\ee
The crucial fact to note here is that the volume dependence comes from assuming that the coupling starts to run from the string scale.
The physics of this is simply that the QCD scale is a function of the scale from which the coupling starts running. However in
string theory with a canonical gravitational action the string scale is itself a function of the moduli, and so the QCD scale itself
is a function of the moduli.

This allows us to add an extra term to the moduli potential, coming from the interaction with matter
\be
\Lambda_{QCD}^{4}\propto \left(\frac{1}{\sqrt{\mathcal{V}} }\right)^{4}=e^{-\sqrt{6}\Phi}.
\ee
The full potential then becomes
\be
V=(1- \alpha\Phi^{3/2}) e^{-\sqrt{\frac{27}{2}}\Phi}+ \epsilon e^{-\sqrt{6}\Phi}+\rho_{0}e^{-\sqrt{6}\Phi},
\label{eq:VFULL}
\ee
where we take
\be
\rho_{0}\propto \Lambda_{QCD}^{4} e^{\sqrt{6} \Phi_{0}}.
\ee

At this point we must note that the assumption that the gauge coupling starts to run from the string scale can be relaxed.
For example, in local models it is not the string scale but instead the winding scale from which couplings start running \cite{Conlon:2009xf}. One can then take
\be
\mu'' \propto \mathcal{V}^{-n},
\ee
which implies $\Lambda_{QCD} \propto \exp(-\sqrt{\frac{-3 n^{2}}{2}}\Phi)$, and 
therefore the term in the potential that parametrises the interaction with matter becomes
\be
\Lambda_{QCD}^{4}\propto e^{-\sqrt{24 n^{2}}\Phi}.
\ee
Some potentially interesting cases include the Kaluza-Klein scale, $M_{KK}=M_{P}/\mathcal{V}^{2/3}$ and the Unification scale  $M_{GUT}=M_{P}/\mathcal{V}^{1/3}$. Throughout the rest of this paper we will study the case where the couplings start to run from the string scale. The results from  Kaluza-Klein our GUT scale running will be essentially the same with the feature that the larger the $n$, the smaller the modulus vev shift and the denser the environment required to destabilise the modulus form its vacuum vev.

In the interests of having a well-defined model, we will assume throughout that the coupling to matter has the form derived above and the potential for the  modulus field is given by Eq. (\ref{eq:VFULL}).

\section{Analysis of the volume modulus potential} \label{sec:analysis}

In this section we analyse the potential for the volume modulus, Eq. (\ref{eq:VFULL}), investigating
how the matter contribution can distort the potential and potentially lead to runaway. This will happen in the region of moduli space where the local energy density is comparable to the combination of the LARGE volume potential plus uplifting term.
For a fixed matter energy density, it is possible to achieve this by tuning the  $\alpha$  parameter in the potential, which is related to the Euler number of the extradimensional manifold.\footnote{This modifies the gravitino mass and gravity-mediated susy breaking scale in the theory.}
For fixed $\alpha$, the same effect occurs as the matter energy density is increased.

This potential tuning process is not completely free. One important constraint to this analysis comes from fifth force experiments. These limit the range of allowed masses for the volume modulus. For gravitational strength fifth force, the modulus mass must lie outside the range $[10^{-17},10^{-2}]$ eV (see e.g. \cite{Adelberger:2003zx}).
We will first analyse the case where the local source is a neutron star and then perform a more generic analysis.

\subsection{Neutron Stars}

We first examine the possibility of shifting the moduli vevs in a neutron star.  We start  by
analysing the case where moduli physics is Planck coupled and then relax this assumption, treating the coupling as an extra free parameter.

We can approximate the potential in the vicinity of the minimum as
\be
V(\Phi) = m^2 (\Phi - \Phi_0)^2 + \Lambda_{QCD}^4 e^{-(\Phi - \Phi_0)/M_X},
\ee
Here $M_X = \lambda M_P$ gives the coupling strength of the modulus - if $\lambda \sim \mc{O}(1)$ then the modulus is Planck coupled,
whereas if $\lambda \ll 1$ then the modulus-matter coupling is stronger than gravitational. $\Phi_0$ is the vacuum expectation
value of the modulus.

From this we see that the shift in the modulus away from its vacuum value $\Phi_0$ is given by
\be
(\Phi - \Phi_0) \simeq \frac{\Lambda_{QCD}^4}{2 m^2 M_X}.
\ee
Recalling that in string theory the high energy couplings are directly related to the vevs of moduli fields, one observes that the fractional shift in a dimensionless coupling  is of order
\be
\label{jabba}
\frac{(\Phi - \Phi_0)}{M_X} \simeq \frac{\Lambda_{QCD}^4}{2 m^2 M_X^2}.
\ee
These expressions are valid for small shifts in the modulus vev: for larger shifts, the global structure of the potential
will become relevant.

\subsubsection{Planck coupled moduli}

In general we expect moduli to be Planck-coupled, and this is true for the volume modulus in the LARGE volume models.
Since $\Lambda_{QCD}^{4}\approx 10^{-80} M_{P}^{4}$, in order to get the minimum inside the compact object to differ from the one outside we are required to work at very large $\Phi$ where the local energy density effects become comparable to the background potential. This in itself does not pose any problem. However the mass of the modulus is given by  Eq. (\ref{eq:masses}) and is also determined by $\Phi$.
For $M_X \sim M_P$ we see from Eq. (\ref{jabba}) that an $\mc{O}(1)$ shift in couplings (equivalently an $\mc{O}(M_P)$ shift in the modulus vev)
requires a modulus mass of $m \sim 10^{-11} \hbox{eV}$ (this corresponds to $\alpha \approx 3.1 \ti 10^{-3}$ and $\Phi \approx 48$).
However this mass value falls within the range that is excluded by consideration of fifth force experiments and in the context of the
large volume models would also require a string scale of around $10 \hbox{keV}$, which is manifestly excluded.

\subsubsection{Strongly coupled moduli}

In a more phenomenological approach one might consider allowing the coupling strength $M_X$ to be a free parameter
rather than tying it to the Planck scale $M_P$, even though in the model considered
the volume modulus with the specific potential, Eq. (\ref{eq:Vfull}), is necessarily Planck coupled.
Moduli in string models can certainly be coupled much more strongly than $M_P$
(in the large volume models the blow-up moduli have matter couplings suppressed by $M_S$ for example).

In this case we require the modulus mass not to be smaller than $10^{-3} \hbox{eV}$ and require an $\mc{O}(1)$ shift in
a coupling. Depending on the precise value of the modulus mass, a numerical analysis now reveals that the interesting range
for $M_{X}$ is between $10^{7}$ and $10^{10} $ GeV, consistent with the estimates of Eq. (\ref{jabba}).
Such strongly coupled moduli are in principle obtainable for models with low string scales. However this does not
really help us, as we expect light moduli with
masses $m \lesssim 1 \hbox{eV}$ and couplings only suppressed by $M_X \sim 10^9 \hbox{GeV}$ to be excluded by
bounds on the cooling of SN1987A by emission of exotic light particles.\footnote{The cooling of SN1987A excludes axion decay constants
$f_a \lesssim 10^{10} \hbox{GeV}$ - see for example \cite{Raffelt}. The precise numbers entering the bound do depend on
the pseudoscalar nature of the axion coupling, and so would be modified for emission of a modulus, but we do not expect the order
of magnitude of the bound to change.}

\begin{figure}[h!]
	\centering
	\begin{minipage}[b]{0.45\linewidth}
	\centering
	\includegraphics[width=\textwidth]{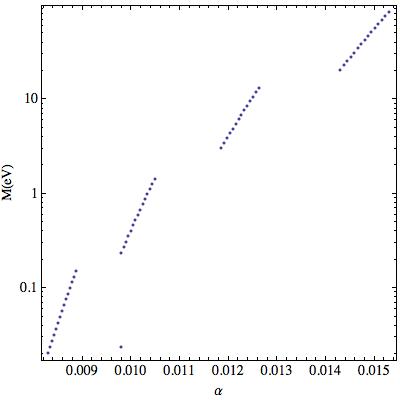}
	\caption{Modulus mass  as a  function of the $\alpha$ parameter for, from left to right, $M_x=10^{10},10^{9},10^{8},10^{7} \hbox{GeV}$}
	\label{fig:mass}
	\end{minipage}
	\hspace{0.5cm}
	\begin{minipage}[b]{0.45\linewidth}
	\centering
	\includegraphics[width=\textwidth]{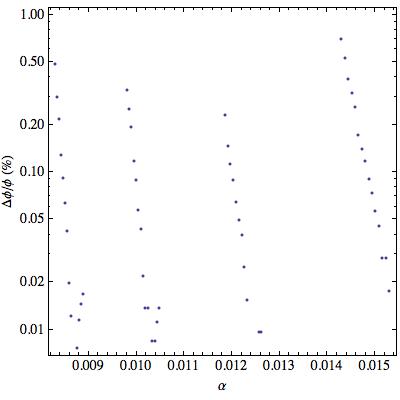}
	\caption{Normalised $\Delta \phi$   as a  function of the $\alpha$ parameter for, from left to right, $M_x=10^{10},10^{9},10^{8},10^{7} \hbox{GeV}$}
	\label{fig:deltaPhi}
	\end{minipage}
	\end{figure}

\subsection{Cosmic strings and other VERY dense objects}\label{cosmString}

We now put neutron stars to one side and consider the typical density that would be required in order to
cause Planckian displacements of moduli. We stress that we here keep
the requirement that moduli physics is Planck coupled, as expected from string theory.
We again aim to find the region of parameter space which allows for the shift of moduli vevs
while remaining  compatible with constraints from fifth-force experiments.

The fundamental reason why neutron star energy densities could not destabilize Planck coupled moduli
is due to the hierarchy $\left(\frac{\Lambda_{QCD}}{M_Pl}\right)^2\approx 10^ {-40}$ in (\ref{jabba}).
It is therefore clear that we would need objects of higher energy density.

\begin{figure}[h]
	\centering
	\begin{minipage}[b]{0.45\linewidth}
	\centering
	\includegraphics[width=\textwidth]{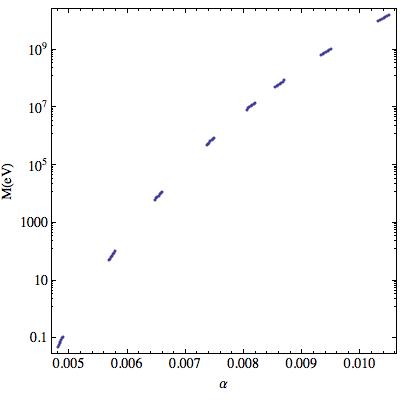}
	\caption{ Modulus mass   as a   function of the $\alpha$ parameter for, from left to right, $\rho=10^{-60},10^{-55},10^{-50},10^{-45},10^{-44},10^{-42},10^{-40},$ $10^{-38}  M_{Pl}^ 4$.}	
	\label{fig:massString}
	\end{minipage}
	\hspace{0.5cm}
	\begin{minipage}[b]{0.45\linewidth}
	\centering
	\includegraphics[width=\textwidth]{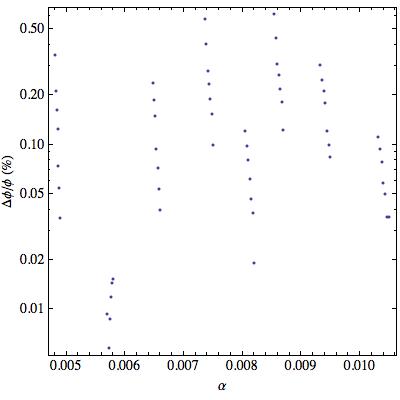}
	\caption{ Normalised $\Delta \phi$   as a   function of the $\alpha$ parameter for, from left to right, $\rho=10^{-60},10^{-55},10^{-50},10^{-45},10^{-44},10^{-42},10^{-40},$ $10^{-38}  M_{Pl}^ 4$.}	
	\label{fig:deltaPhiString}
	\end{minipage}
	\end{figure}

In what follows we perform a numerical scan for objects of energy densities $\rho \in [10^{-60}, 10^{-44}] M_{Pl}^ 4$.
The results of the numerical study are shown in Figs. \ref{fig:massString} and \ref{fig:deltaPhiString}.
We require an $\mc{O}(0.1)$ fractional displacement of the moduli from its vacuum value in a dense background.
We plot in Fig. \ref{fig:massString} the vacuum mass of the modulus for which this can be attained, for several
different values of the background density. One sees that as the
local perturbation becomes less dense, the region of parameter space where a deviation of order $1\%$ in the modulus vev is
attained corresponds to a region where the mass for this modulus is smaller.
Keeping in mind that the fifth force lower limit for the mass is around $10^{-2} \hbox{eV}$ we conclude that
the minimum energy density of an object capable of generating regions of different vacuum while still being
compatible with fifth force constraints is $\rho \approx 10^{-60} M_{Pl}^{4}$.\\
Figure  \ref{fig:deltaPhiString} shows the fractional shift in the modulus vev for various energy densities.

Having performed this generic analysis we should now consider where such densities could come from. We first note that
any such object has to be \emph{extremely} dense, about 20 orders of magnitude denser than neutron stars. For static objects
a couple of possibilities present themselves. First, there are GUT cosmic strings.
These topological defects are remnants of the GUT breaking which may have happened early in the history of the universe.
Their mass per unit length $\mu$ is related to the scale of the breaking of the symmetry that generates them $\sigma$ by $\mu \propto \sigma^ 2$. In the case of GUT strings, $\sigma\approx 10^{16} \hbox{GeV}$ and therefore $\mu \approx 10^{32} \hbox{GeV}^2$. For cosmological purposes the strings are taken to be one dimensional objects,  since they have to be either closed (unstable) or infinite. This is a simplification and from the theoretical point of view  they are known to have a finite radius of the order of the correlation length of the field that spontaneously breaks the GUT.

Secondly, one could imagine dark sector analogues of neutron stars - compact bound objects held together by degeneracy pressure in the
same way as neutron stars are, in the case that there existed a dark analogue of QCD with a confinement scale $\Lambda_{dark} \gtrsim
1 \hbox{TeV}$. To avoid black hole formation such objects would need to be highly compact with rather small radii.

However, there is no strong reason to think either of these two objects exist, although we will further
discuss the latter case in Sec. \ref{sec:statCond}. One
case which does exist and where highly energy-dense regions are expected is that of
singularities arising either cosmologically or through black hole formation. This will be analysed in the next sections.

\subsection{Cosmological singularities}\label{sec:CosmSing}

The fact that near cosmological singularities energy densities may be arbitrarily high
makes them good candidates for the study of a coupled modulus/matter system.
This means that not only is it possible to find destabilise a minimum for the moduli potential but also that  decompactification might be possible.

The model we consider is that of a closed, matter dominated FRW universe.
The overall spacetime structure is $\mathcal{R}^{+}\times S^3\times CY^6$.
As before we assume that the matter is baryon-like, with a mass depending on the volume modulus as in Eq. (\ref{proton}).
We take the modulus scalar field potential to be given as before by the
LARGE volume potential plus the up lifting term.
We assume the volume modulus is initially at the minimum and unexcited, so
that $\Omega_{\Phi,init} \equiv (\rho_\Phi /\rho_{tot})_{init} = 0$ while $\Omega_{matter, init}=1$.
The matter-moduli coupling
comes from the field dependent mass in Eq. (\ref{proton}), and takes the form
\be
\rho(a,\Phi)=\frac{\rho_0}{a^ 3}e^{-\sqrt{6}(\Phi-\Phi_0)}.
\label{eq:rho}
\ee

The evolution of the system is determined, as usual, by the Friedmann equation
\be
\left(\frac{\dot{a}}{a}\right)^ 2=\frac{1}{3}\left(\frac{\dot{\Phi}^ 2}{2}+V(\Phi)+\rho(a,\phi)\right)-\frac{\kappa}{a^2},
\label{eq:Friedman}
\ee
and the Klein-Gordon equation for a homogeneous and isotropic field
\be
\ddot{\Phi}=3 \frac{\dot{a}}{a}\dot{\Phi}+(V(a,\Phi)+\rho(a,\Phi))_{,\Phi}.
\label{eq:KG}
\ee

\begin{figure}[h]
\centering
\begin{minipage}[b]{0.45\linewidth}
\centering
\includegraphics[width=\textwidth]{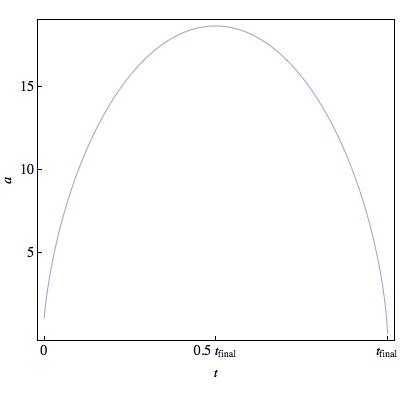}
\caption{Time evolution of the scale factor in a closed, matter dominated FRW universe.}
\label{fig:a}
\end{minipage}
\hspace{0.5cm}
\begin{minipage}[b]{0.45\linewidth}
\centering
\includegraphics[width=\textwidth]{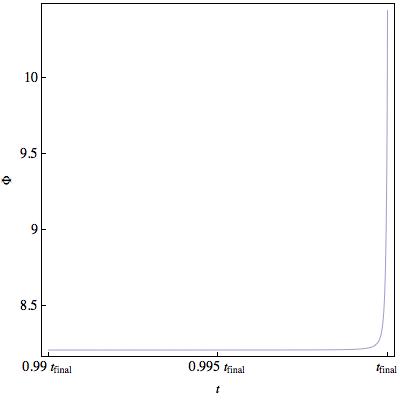}
\caption{Time evolution of the modulus in a closed, matter dominated FRW universe.}
\label{fig:phi}
\end{minipage}
\end{figure}

To study the coupled modulus-matter dynamics we solve the equations of motion
numerically using a modified version of \cite{Kallosh:2004rs}.
 As expected the initial evolution of this universe is the same as a regular closed matter dominated universe: the scale factor
 grows to a maximum before the universe will start collapsing, with the scale factor shrinking to zero.
 This behaviour is shown in Fig.  \ref{fig:phi}.
During most of the collapse, the system will not show any departure from the behaviour of a closed, matter dominated FRW universe. It is only when the scale factor becomes sufficiently small (let's call it $\overline{a}$), that the term in Eq. (\ref{eq:rho}) will become large
 enough to modify the structure of the modulus potential, Eq. (\ref{eq:VFULL}), and play a significant role in the evolution of the field.
This happens, to first order, when the matter contribution to the potential at the minimum is comparable to the potential plus matter at the maximum, i.e.
\be
V(\Phi)+\rho(a,\Phi)|_{min_0}=V(\Phi)+\rho(a,\Phi)|_{max_0}.
\label{eq:decompCond}
\ee
 Note that  $min_0$ and $max_0$ denote the minimum and maximum of the LARGE volume plus uplifting potential (which differ from the extrema of the full potential). One can get an semi analytical estimate for $\overline{a}$, by solving Eq. (\ref{eq:decompCond}), finding
\be
\overline{a}=\left( \frac{\rho_0}{V(\Phi_{max_0})}(e^{-\sqrt{6}\Phi_{min_0}}-e^{-\sqrt{6}\Phi_{max_0}})\right)^ {1/3}.
\label{eq:abar}
\ee
This estimate can be refined by expanding $a_c=\overline{a}+\delta a$ and solving the condition
\be
V(\Phi)+\rho(\overline{a}=\delta a,\Phi)|_{min}=V(\Phi)+\rho(\overline{a}+\delta a,\Phi)|_{max},
\label{eq:decompCond2}
\ee
where $min$ and $max$ denote the position of the minimum and the maximum of the full potential, i.e. including matter contribution, when $a=\overline{a}$. Solving Eq.(\ref{eq:decompCond2}) one finds
\be
a_c\equiv\overline{a}+\delta a =\left( \frac{\rho_0}{V(\Phi_{max})-V(\Phi_{min})}(e^{-\sqrt{6}\Phi_{min}}-e^{-\sqrt{6}\Phi_{max}})\right)^ {1/3},
\ee
which is in good agreement with numerical estimates.

As the scale factor approaches $a_{c}$, the modulus expectation value begins to shift and the barrier to decompactification
decreases. After a short period of evolution, the term (\ref{eq:rho}) becomes the dominant term in the potential and there is no
obstacle to prevent the field $\Phi$ from rolling towards infinity. The evolution of $\Phi$ is shown in Fig.  \ref{fig:phi} and we
see the sharp increase in $\Phi$ beyond a critical time. This final stage of the evolution sees a rapid runaway of $\Phi$ to
infinity while the scale factor continues to shrink to zero. The energy density of the universe will quickly be dominated by the kinetic energy of the volume modulus, with the dust and potential contributions becoming subdominant. The evolution of the system is depicted in Figs. \ref{fig:a}, \ref{fig:phi} and 9.

\begin{figure}[h!]
\centering
	\includegraphics[width=0.45 \textwidth]{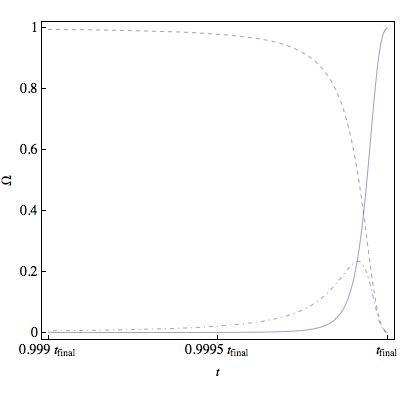}
	\label{fig:omega}
	\caption{Late time evolution of the energy densities of matter and scalar field (kinetic and potential). Dashed line $\Omega_{dust}$, dashed-dotted line $\Omega_{potential}$, full line $\Omega_{kinetic}$. }
	\end{figure}

In this last region it is possible to solve the equations of motion analytically. The equations of motion are
\bea
\frac{d}{dt} \left( a^3 \dot{\Phi} \right) & = & - a^3 \left( \frac{\partial V}{\partial \Phi} \right), \\
\label{eq412}
\frac{d}{dt} H & = & - \frac{\dot{\Phi}^2}{2} + \frac{\kappa}{a^2}.
\eea
Once $\Phi$ starts to runaway, the potential vanishes exponentially fast and ceases to be a significant contribution.
In this asymptotic regime we can then solve these equations by
\bea
\label{sol1}
\Phi(t) & = & \Phi_0 - \sqrt{\frac{2}{3}} \ln \left( t_0 - t \right), \nn \\
a(t) & = & \left( \frac{3}{2} \right)^{1/6} (t_0 - t)^{1/3}.
\eea
From the solution we see that we can self-consistently neglect the curvature term in Eq. (\ref{eq412}) for $\frac{t_0 - t}{t} \ll 1$.
We can relate the canonically normalised field $\Phi(t)$ to the volume of the compact space using Eq. (\ref{eq:Phi}), to find that the
compact volume $\mc{V}$ evolves as
\be
\label{vol1}
\mc{V} = \frac{\mc{V}_0}{t_0 - t},
\ee
and so diverges as $t \to t_0$.

In the limit as $t_0 \to t$ this therefore describes a universe where
\begin{enumerate}
\item
The 4-dimensional scale factor shrinks to zero size as $(t_0 - t)^{1/3}$.
\item
The volume of the six compact dimensions is first destabilised before diverging
as $(t_0 - t)^{-1}$.
\item
There is a singularity at $t = t_0$, at which the 4-dimensional scale factor is formally zero and
the 6-dimensional volume is formally infinite.
\end{enumerate}
This represents a universe which bounces, albeit heterogeneously: as the three spatial dimensions collapse, the extra six dimensions
expand and reach infinite volume in finite time.

Let us consider the validity of a 4-dimensional effective field theory treatment. From the solution for
$\Phi(t)$ in Eq. (\ref{sol1}) it is clear that the (kinetic) energy density of the $\Phi$ field diverges in finite time,
\be
\rho_{\Phi} = \half \dot{\Phi}^2 = \frac{1}{3 (t_0 - t)^2}.
\ee
Furthermore as decompactification occurs and the extra-dimensional volume increases, the 4-dimensional string scale
decreases.
\be
M^4_{s,4d} = \frac{M^4_P}{\mc{V}^2} = M_P^4 (t_0 - t)^2 e^{-\sqrt{6} \Phi_0}.
\ee
Consequently, independent of initial conditions, as $t \to t_0$ the system will evolve to a state where the kinetic energy
density in the field $\Phi$ is greater than the apparent cutoff of a 4d effective field theory.

Fortunately it is easy to understand what is happening from a higher dimensional perspective. Once the field $\Phi$
gets over its decompactification barrier, its potential soon becomes negligible - we can see in Fig. 9
how the kinetic energy of $\Phi$ is the dominant contributor to the energy density. Although we have written the problem
in the language of 4-dimensional effective field theory, $\Phi$ is originally the volume modulus of the extra dimensions.
Neglecting the potential energy, the system is then fundamentally that of 10-dimensional general relativity. The above solution
then corresponds to a Kasner solution dimensionally reduced to 4 dimensions.

It is not immediately clear that the above numbers are consistent with a Kasner solution. Recall that the Kasner solution is
\be
\label{kasner}
ds^2 = - dt^2 + \sum_i t^{2 p_i} dx_i^2,
\ee
with $\sum p_i = 1$ and $\sum p_i^2 = 1$. For a $1+3+6$ dimensional Kasner solution, the allowed exponents are (using $p_3$ to denote
the 3-dimensional growth and $p_6$ for the six-dimensional growth)
\be
p_3 = -\frac{1}{3}, p_6 = \frac{1}{3}, \qquad p_3 = \frac{5}{9}, p_6 = \frac{-1}{9}.
\ee
It is clear that the evolution of the scale factor in (\ref{sol1}) and the volume in
(\ref{vol1}) do not fit these conditions. However, note that the metrics are different: the metric used in the Kasner
solution (\ref{kasner}) is the 10-dimensional string frame metric, whereas the 4-dimensional metric for which Eqs.
(\ref{sol1}) and (\ref{vol1}) applies is a dimensionally reduced metric that is related to the 10-dimensional metric by factors
of the internal volume. It is then expected that the 4-dimensional scale factor does not have the Kasner exponent appropriate for a
10-dimensional (1+3+6) solution.

There is one striking feature about this behaviour. The initial conditions (a closed matter-dominated FRW universe) were unexceptional.
However these conditions unavoidably evolve to give dynamic super-inflationary behaviour of the compact dimensions, which in theory reach
infinite volume in finite time. Furthermore, this behaviour commences in the region controlled by effective field theory, where a
4d description is valid. In practice, the evolution of a Kasner solution should break down as the contracting dimensions approach the 10d string scale, regulating the infinity.
This super-inflationary behaviour cannot be prevented - growth in the 4d energy density is
a necessary consequence of a spacetime crunch - and this
energy density must always eventually overcome the barrier to decompactification.

As formulated the dynamics have started with 3 large and 6 compact dimensions, ending with 6 large and 3 compact dimensions. However the physics is such that there is no reason not to reverse the process, and imagine starting with 6 large and 3 compact dimensions and ending with 6 compact and 3 large dimensions. In effect the universe bounces within effective field theory, which is achieved by the bounce occuring in different dimensions to the collapse: the collapse of certain dimensions triggers the expansion of others. This physics has some
similarities to pre-Big Bang cosmology \cite{Veneziano1991, hepth9211021, hepth0207130}. This was formulated using the heterotic dilaton
and the $\mc{O}(d,d)$ symmetries of toroidal compactification.
It would be interesting to make the connections more precise
and see whether this super-inflationary growth of the compact dimensions is able to mimic some of the physics of conventional inflation.
We leave this to future work.

\subsection{A static solution}\label{sec:statCond}

In this section we study spherically symmetric configurations and analyse the resulting modulus profile.
The models considered in this section are similar to the `dark stars' mentioned in Sec. \ref{cosmString}: we look for stable
solutions of matter coupled to a modulus field, protected against gravitational collapse.
In Sec. \ref{subsub} below we allow for dynamical evolution of the profile and study gravitational collapse of dust balls.

Consider a spherical ball of dust of radius $R$. The dust has energy density $\rho$ but is pressureless ($P = 0$).
Let the line element inside the dust ball be curved FRW:
\be
\label{FRWFRW}
ds^{2}=-dt^{2}+a^{2}\left( \frac{dr^{2}}{1-\kappa r^{2}}+r^{2}(d\theta^{2}+ \sin^{2}\theta d\phi^{2}) \right),
\ee
where $a$ is the scale factor and $\kappa$ is the spacial curvature.  By Birkoff's theorem, the spacetime for $r>R$ is the Schwarzschild solution:
\be
ds^{2}=-\left(1-\frac{2M}{r}\right)dt^{2}+ \frac{dr^{2}}{1-\frac{2M}{r}}+r^{2}(d\theta^{2}+\sin^{2}\theta d\phi^{2}),
\ee
where M is the total mass in the region $r<R$.

For arbitrary values of $\rho$ and $P$, the equations of motion for the scale factor $a(t)$ in Eq. (\ref{FRWFRW}) are:
\bea
\label{eq:FRW1}
\frac{\ddot{a}}{a}=-(\rho+3 p),\\
\label{eq:FRW2}
\left(\frac{\dot{a}}{a}\right)^{2}=\frac{\rho}{3}-\frac{\kappa}{a^{2}}.
\eea
One can find a static solution ($\dot{a}=\ddot{a}=0$) of Eqs. (\ref{eq:FRW1}), (\ref{eq:FRW2}) by taking
\bea
\label{eq:StatCond1}
\kappa=\rho/3,\\
\label{eq:StatCond2}
\rho=-3 p,
\eea
where we have set $a=1$.  This is the Einstein Static Universe.

In the spirit of the previous sections, let the mass of the dust particles be a function of the modulus field $\phi$. The dust energy density is therefore given by:
\be
\rho_{dust} =\rho_{0} e^{-\sqrt{6}\phi},
\ee
which represents a source for the volume modulus potential in the region $r<R$. The dust has zero pressure but there is a pressure contribution from the vacuum energy of the modulus field as it is displaced from its minimum.

The simplest modulus profile compatible with the equations of motion is obtained by considering three distinct regions. The first region is $r<R$, in which the field is at the minimum of the effective potential $V_{eff}=V_{\phi}+\rho_{dust}$. For $r>R+1/m_{\phi}$ the field is at its vacuum minimum. In the transition region $R<r<R+1/m_{\phi}$ the field interpolates smoothly between the two distinct minima.

The volume modulus profile described above implies that inside the dust ball,
\bea
\rho& = & \rho_{dust}+V_{\phi}, \\
p & = & -V_{\phi},
\eea
where these are evaluated at the minimum of the combined potential $V_{dust} + V_{\phi}$. Although there is no pressure from the dust,
the displacement of scalar fields from the minimum leads to a contribution to vacuum energy.
One then finds that Eqs. (\ref{eq:StatCond1}) and  (\ref{eq:StatCond2}) become
\bea
\kappa=\frac{\rho_{dust}+V_\phi}{3},\\
\label{eq:StatCondFINAL}
\rho_{dust}-2 V_{\phi}=0.
\eea
Since one can treat the spatial curvature as a free parameter, the condition for existence of a static solution for the region $r<R$ reduces to Eq. (\ref{eq:StatCondFINAL}). As long as the combined potential $V = V_{dust} + V_{\phi}$ exhibits a minimum where $V_{dust} = 2V_{\phi}$,
then a static solution will exist.
There are two possibilities of tuning the system to generate a solution to Eq. (\ref{eq:StatCondFINAL}):
one may either tune the dust density $\rho_0$ or the parameter $\alpha$ in the LARGE volume potential.
There is then effectively a 1-parameter set of solutions parametrised by the density of the interior.
There are various constraints on this set, for example by
limits on the volume modulus mass coming from fifth force experiments, as discussed in Sec. \ref{sec:analysis}.

In Table \ref{tab:StaticSol} we display four different static solutions. These were obtained by fixing $\alpha$ and then tuning the density $\rho$.
It is interesting to study the properties of the dust distribution that sources the  nontrivial volume modulus profile, in particular its mass and radius. Given that the density is uniform and fixed by Eq. (\ref{eq:StatCondFINAL}), the mass will be given by $
M=\frac{4\pi}{3}\rho R^{3},$ where $R$ is the radius of the spherical dust distribution. One may write the radius in terms of the Schwarzschild radius as $R=\xi R_{sch}$, where $\xi>1$. Since, by definition, $R_{sch}=2GM$ one finds
\be
R_{sch}=\sqrt{\frac{3}{8 \pi G \rho \xi^{3}}},
\ee
this implies that
\be
R=\sqrt{\frac{3}{8 \pi G \rho \xi}}.
\ee

One then concludes that the properties of the dust ball are completely determined by $\xi$ since $\rho$ is fixed by requiring a static solution. Note that the smaller $\xi$ the larger the star's radius and mass. In Table \ref{tab:StaticSol} we display the radii and masses for the four cases under study considering $\xi=1.1$.

\begin{table}[h]
\begin{center}
\begin{tabular}{c|c|c|c|c}
$\alpha$ & $m_{\Phi} (M_{p})$&$\rho (M_p^4)$&$R (M_p^{-1})$& $M (M_{p})$\\
\hline
\hline
$0.05$&$1.42\times10^{-7}$&$ 5.39278\times 10^{-7}$&$ 449$ & $204$\\
$0.01$&$ 4.45\times10^{-19}$&$ 5.36521\times 10^{-15}$&$ 4.49\times10^{6}$&$2\times 10^{6}$\\
$0.005$&$2.80\times10^{-29}$&$ 6.288865\times 10^{-22}$&$ 1.3\times 10^{10}$ & $6\times 10^{9}$\\
$0.001$&$ 5.16\times 10^{-82}$&$ 2.15993\times 10^{-57}$&$ 7\times 10^{27}$&$3\times10^{27}$\\
\end{tabular}
\end{center}
\caption{Mass of the canonically normalised volume modulus, density, radius and total mass of the dust sphere, as functions of the $\alpha$ parameter in the LARGE volume potential. Radius and total mass computed assuming $\xi=1.1$.}
\label{tab:StaticSol}
\end{table}%

The numerical results in table \ref{tab:StaticSol} reveal that there is a very large hierarchy both in radius and mass between the various cases studied here. This exemplifies the issue raised in section 4.2: for masses of $\Phi$ greater than that allowed by fifth force constrains, namely
$m_{\Phi} \gtrsim 10^{-30} M_P$, the size of such objects is extremely small ($R \lesssim 10^{-23} \hbox{cm}$).
The corresponding mass is $M \lesssim 1 \hbox{kg}$. While they could in principle form part of dark matter, it is hard to see how
interesting physics can be extracted from them. The local density of dark matter objects with $\hbox{kg}$ masses is not larger than
$10^{-21} m^{-3}$ and so objects would be both unobservable and undetectable.

\subsection{A dynamic solution: black hole formation}
\label{subsub}

In this section we generalise the analysis of Sec.\ref{sec:statCond} to allow for dynamical evolution of the system.
While Sec. \ref{sec:statCond} was restricted to static solutions, here we look for collapsing solutions of the coupled matter-modulus system.
If one lets the gravitational collapse last for long enough, the final state will be a Schwarzschild black hole.
However before one reaches the singularity, the local density becomes arbitrarily large, which could be sufficient to destabilise the modulus.

Here our initial conditions are a large, dilute dust ball which we allow to collapse towards a black hole.
As before we assume that the spacetime is given by a curved FRW universe smoothly connected at the surface of the dust sphere ($r=R$) to a Schwarzschild solution. In contrast to Sec. \ref{sec:statCond}, we now use the Friedmann equation to obtain dynamical solutions of the scale factor and of the energy density for $r<R$.

The profile for the volume modulus is obtained by considering:
\begin{equation}
 \phi = \left\{ \begin{array}{ll}
	\ddot{\phi}+ 3 \frac{\dot{a}}{a}\dot{\phi} +V_{,\phi}=0& ,r\in[0,R]\\
  	(1-2M/r)\phi''+(\frac{2}{r}(1+2M/r)+2M/r^{2})\phi'-V_{,\phi}=0 & ,r\in[R,R+m_{\phi}^{-1}]\\
 	\phi_{\infty} & ,r\in[R+m_{\phi}^{-1},\infty]
\end{array} \right.
\label{eq:profile}
 \end{equation}
where  $'\equiv \frac{\partial}{\partial r}$, $\dot{}\equiv  \frac{\partial}{\partial t}$, M is the mass sourcing the Schwarzschild geometry and $\phi_{\infty}$ is the value of the field in the minimum of its vacuum potential.

The time dependence of the system arises in two different ways. Inside the dust distribution one has a homogeneous and isotropic positively curved universe, which is in general dynamic. This causes the scale factor and the volume modulus to be functions of time. For $r>R$ we assume that the time dependence comes only from the time variation of the radius of the dust ball (which can be traced to the time variation of the scale factor).
In particular, we assume that the scalar field profile adjusts instantaneously to the change in dust density, i.e. that
the characteristic time scale of the gravitational collapse is much larger than the typical timescale of the variation of the scalar field.

One must clarify what is meant by $M$ in Eq. (\ref{eq:profile}). Naively one would expect the mass of the dust ball to be given by the volume integral of the dust energy density. However there are also non negligible contributions from the potential and kinetic energies of the volume modulus. These are, for most of the evolution of the system, one order of magnitude smaller than the dust contribution but become more significant as the system evolves. Taking this effect into consideration, the mass sourcing the Schwarzschild geometry is

\be
M=\int d^{3}x \sqrt{g} \left(\rho_{dust}+\rho_{V(\phi)}+\rho_{K(\phi)}\right).
\label{eq:2M}
\ee
\begin{figure}[h]
\begin{center}
\includegraphics[width=0.5\textwidth]{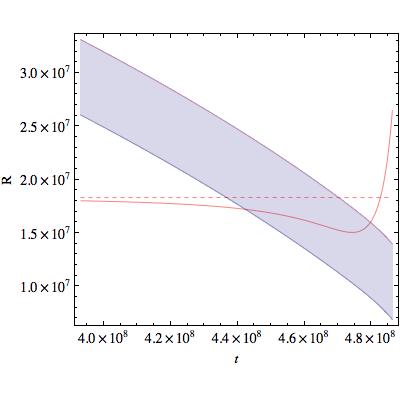}
\caption{Comparison between the naive estimate for 2M (dashed line)  and the result of Eq. (4.33) (solid line).
The shaded area represents the region where the field varies between the value inside the dust ball and the vacuum minimum.}
\label{fig:2Mtot}
\end{center}
\end{figure}
In Fig.  \ref{fig:2Mtot} we compare Eq. (\ref{eq:2M}) with the naive estimate. The deviation between the two estimates is negligible throughout most of the evolution but increases with time, diverging at the end. This divergence is due to the fact that when the density reaches a critical value, the volume modulus potential no longer exhibits a minimum and decompactification happens. This means that the modulus starts to roll and its kinetic energy dominates the energy density inside the dust ball, in a similar way to the behaviour described in Sec. 4.3 for cosmological singularities.

\begin{figure}[h]
\begin{center}
\includegraphics[width=0.9\textwidth]{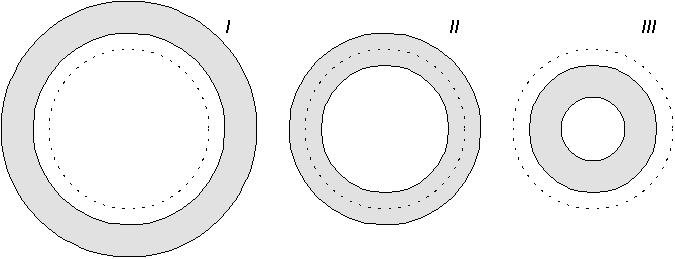}
\caption{Sketch of the black hole formation process. The dashed line depicts the Schwarzschild  horizon and the shaded area depicts the region $R<r<R+1/m_{\phi}$ where the field rolls between its value inside and outside the dust ball. }
\label{fig:BHSketch}
\end{center}
\end{figure}

The time evolution of the system is depicted in Fig.  \ref{fig:BHSketch}. We start with a spherical distribution of radius $r>2M$, regime I. In this regime the whole transition region (shaded area) is outside the horizon and the deviation of the volume modulus from its vacuum minimum
 could in principle be observable. As the gravitational collapse evolves a black hole will form and subsequently the transition region will start to fall inside the horizon, this is regime II. In this regime, it is still in principle possible to observe the consequences of the matter modulus interaction since part of the transition region lies outside the horizon. When the whole of the transition region is veiled by the horizon, regime III, an observer sitting outside the black hole will not be able to measure the  nontrivial profile of the volume modulus.

\begin{figure}[h]
\begin{center}
\includegraphics[width=0.5\textwidth]{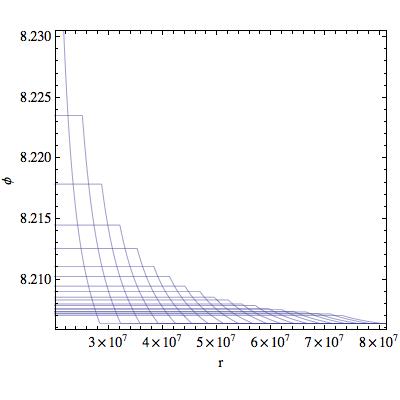}
\caption{Time evolution of the volume modulus profile in regime I.}
\label{fig:profile}
\end{center}
\end{figure}

The time evolution of the volume modulus profile in regime I is shown in Fig.  \ref{fig:profile}.  As the system collapses and the density increases, the volume modulus vev inside the dust ball increases and the transition region where the field is allowed to vary moves to the smaller radius region. The evolution of the field profile in regimes II and III is qualitatively similar to the one just described.\footnote{In regime II one must impose finiteness and continuity of the solution at the horizon}


\section{Discussion}

The purpose of this paper has been to consider the interaction of moduli and matter fields, and specifically to analyse the
circumstances under which moduli fields can be destabilised from their vevs by dense concentrations of matter.
We have described the origin of moduli/matter couplings and the assumed form of the moduli potential.
We have tried to consider `honest' values for parameters such as coupling strengths and moduli masses, enforcing the consistency
constraints that emerge from string compactifications.

Our results have both negative and positive elements. On the negative side, it is not feasible to destabilise moduli
through even the densest astrophysical environments. We considered neutron stars as the densest known form of matter, and found that
Planck coupled moduli could only be destabilised if their masses were deep in the region excluded by fifth force constraints.
If the coupling is relaxed from Planck strength, then the values of the coupling for which destabilisation is possible
are excluded by the stellar cooling constraints that require $f_a < 10^9 \hbox{GeV}$. This only leaves very exotic
cases such as cosmic string and hypothetical dark analogues of neutron stars.

On the positive side, our results show that the modulus/matter coupling will play a significant role in gravitational collapse.
The cosmological collapse of a matter-dominated FRW universe was shown to lead to a super-inflationary decompactification of
the internal dimensions as the volume modulus is destabilised. This super-inflationary decompactification leads to
the internal dimensions reaching infinite volume at the same time as the external scale factor vanishes.
It would be interesting to see whether this super-inflationary decompactification can mimic some of the features
of conventional inflation and we leave this for future work.
In a similar vein we also saw that the process of black hole formation will lead to decompactification
during the period of collapse.

\subsection*{Acknowledgments}

JC is supported by a Royal Society University Research Fellowship and by Balliol College, Oxford.
FGP is suported by Funda\c{c}\~{a}o para a Ci\^{e}ncia e a Tecnologia (Portugal) through the grant SFRH/BD/35756/2007.

\bibliographystyle{JHEP}

\end{document}